\documentclass{ieeeaccess}

\usepackage{cite}
\usepackage{amsmath,amssymb,amsfonts}
\usepackage{algorithmic}
\usepackage{graphicx}
\usepackage{textcomp}
\usepackage{epsfig}
\usepackage{endnotes}
\usepackage{listings}
\usepackage{graphicx}
\usepackage{subcaption}
\usepackage{url}
\usepackage{multicol}
\usepackage{enumitem}
\usepackage{supertabular}
\usepackage[T1]{fontenc}
\usepackage{hyperref}
\usepackage{graphicx}
\usepackage{epsf}
\usepackage{fancyvrb}
\usepackage{hyphenat}
\usepackage{subcaption}
\usepackage{multirow}
\usepackage{url}
\usepackage{outlines}
\usepackage{tablefootnote}
\usepackage{xifthen}
\usepackage{pifont}
\usepackage{comment}
\usepackage{balance}
\usepackage{soul}

\newcommand{\cmark}{\ding{51}}
\newcommand{\xmark}{\ding{55}}

\newlength\bubblesize
\newcommand{\yes}{\cmark}
\newcommand{\no}{\xmark}
\newcommand{\na}{---}
\newcommand{\sorta}{\ding{109}}

\def\rot{\rotatebox}

\def\BibTeX{{\rm B\kern-.05em{\sc i\kern-.025em b}\kern-.08em
    T\kern-.1667em\lower.7ex\hbox{E}\kern-.125emX}}
\begin{document}
%\history{Date of publication xxxx 00, 0000, date of current version xxxx 00, 0000.}
%\doi{10.1109/ACCESS.2017.DOI}
\doi{10.1109/ACCESS.2021.3104854}

\title{Review of the Security of Backward-Compatible Automotive Inter-ECU Communication}
\author{\uppercase{Chandra Sharma}\authorrefmark{1},
\uppercase{Samuel Moylan\authorrefmark{2},
Eugene Y. Vasserman}\authorrefmark{1}, \IEEEmembership{Member, IEEE}, and \uppercase{George T. Amariucai}\authorrefmark{1}, \IEEEmembership{Member, IEEE}}
\address[1]{Department of Computer Science, Kansas State University, Manhattan, KS 66506 (e-mail: \{ch1ndra, eyv, amariucai\}@ksu.edu)}
\address[2]{Work performed while at the Department of Computer Science, Kansas State University, Manhattan, KS 66506 (e-mail: smoylan22@ksu.edu)}

\markboth
{Sharma \headeretal: Review of the Security of Automotive Communication}
{Sharma \headeretal: Review of the Security of Automotive Communication}

\corresp{Corresponding author: Chandra Sharma (e-mail: ch1ndra@ksu.edu).}

\begin{abstract}
Advanced electronic units inside modern vehicles have enhanced the driving experience, but also introduced a myriad of security problems due to the inherent limitations of the internal communication protocol. In the last two decades, a number of security threats have been identified and accordingly, security measures have been proposed. While a large body of research on the vehicular security domain is focused on exposing vulnerabilities and proposing counter measures, there is an apparent paucity of research aimed at reviewing existing works on automotive security and at extracting insights. This paper provides a systematic review of security threats and countermeasures for the ubiquitous CAN bus communication protocol. It further exposes the limitations of the existing security measures, and discusses a seemingly-overlooked, simple, cost-effective and incrementally deployable solution which can provide a reasonable defense against a major class of packet injection attacks and many denial of service attacks.
\end{abstract}


\titlepgskip=-15pt

\maketitle

\section{Introduction}
The increasing sophistication of electronic components in modern vehicles has made driving more pleasant, comfortable and, in many cases, safer. The inter-connectivity of the electronic sensors and actuators, and their configurability, helps fine-tune the driving experience, leading in turn to increasing the prevalence and sophistication of these components. Over the years, user-installable data and firmware updates have been introduced, requiring \textit{extra}-vehicular connectivity over wireless protocols.

At the center of most \textit{intra}-vehicular communication lies the CAN bus which connects the electronic control units (ECUs) inside the vehicle.
The CAN specification was developed to meet the real-time communication needs of a vehicle \cite{farsi1999overview, voss2008comprehensible, di2012understanding}, without significant concern for security. Unfortunately, as vehicle manufacturers started adding remote interfaces to ECUs while still following the CAN $2.0$ standard (which never had a security-focused revision), the inherent security limitations started to become an issue. Over the years, numerous critical CAN bus vulnerabilities have been identified.

In the vehicle security domain, there are mainly two (not mutually exclusive) classes of research: the first class focuses on identifying security threats (vulnerabilities, attack surfaces and exploits) and the second class focuses on proposing security measures. One of the earliest works in identifying the security threats to the automotive bus systems, dating back to 2004, is that of Wolf, Weimerskirch, and Paar \cite{wolf2004security}. For the most part, their work covers potential incentives for prospective attackers to hack into the bus, the potential access points that can be used and the general security measures that can be employed to protect access to the bus. Their work, however, fails to provide a practical example of an attack on the bus, nor does it provide concrete implementation details of the discussed security measures. A few years later, in 2010, Koscher et al. \cite{koscher2010experimental} investigate inherent weaknesses in the CAN protocol, look into flaws in the real-world implementation of the protocol (for instance, deviations from the standard) and perform practical attacks on the bus. In 2013, Miller and Valasek open up the true range of automotive attack possibilities. Their work in \cite{miller2013adventures} covers a broad range of attacks leading to the control of different vehicle functionalities, such as braking, steering and acceleration, through physical access to the bus. Similarly, attacks that can be carried out remotely are discussed in \cite{miller2015remote}. 

Somewhat parallel to the discovery of security threats to the CAN bus, a different body of work started to look into proposing security solutions to the identified threats. Claimed by the authors to be the first efficient data authentication scheme for the automotive network, \cite{nilsson2008efficient} is centered on a delayed message authentication based on compound message authentication codes. Similarly, \cite{hoppe2008security} covers intrusion detection techniques based on three detection patterns: \textit{increased message frequency}, \textit{obvious use of message IDs} and \textit{low level communication characteristics} to detect potential attacks on the bus. These two works lay the foundation for many other authentication-based and intrusion detection-based security measures.

Besides the two classes of active research, there is a great need for a third class of research that surveys existing works on automotive security and provides insights. To the best of our knowledge, the literature contains only a handful of papers~\cite{kleberger2011security, studnia2013survey, miller2014survey, avatefipour2018state, liu2017vehicle} that provide an overview, albeit not an extensive one, of security threats to the CAN bus and countermeasures to protect against them. In \cite{kleberger2011security}, the authors survey and identify the underlying security problems of the in-vehicle network. They further review architectural security features proposed by other researchers and discuss the deployment of honeypots and intrusion detection systems that constitute proposed security measures against potential attacks. {Similarly, \mbox{\cite{studnia2013survey}} surveys some of the common attack vectors, both local and remote, and explores the external and internal protection measures to secure the vehicular communication bus.} In \cite{miller2014survey}, the authors discuss a wide class of remote attack surfaces and vulnerable cyber-physical systems and share insights on measures that can be taken to protect against remote attacks. {The work in \mbox{\cite{liu2017vehicle}} explores the vulnerabilities of the in-vehicle network, outlines different attack methodologies and classifies some of the existing countermeasures. Likewise, in \mbox{\cite{avatefipour2018state}}, the authors explore the security limitations of the CAN bus and cover some security measures that researchers have proposed over the years. They also look into some potential attack scenarios and provide the CERT classification of these attack scenarios.}

In a more recent work, Dibaei et al. \mbox{\cite{dibaei2020attacks}} review the vulnerabilities, attacks and defenses on intelligent connected vehicles. They discuss the architecture for intelligent vehicles and highlight the security requirements. They further classify various security attacks and explore existing defenses against those attacks. While their work covers broad security threats and mitigations primarily focusing on the Vehicle-to-Vehicle (V2V) and Vehicle-to-Infrastructure (V2I) communication, it fails to scrutinize subtle security aspects of intra-vehicular communication.

{In this paper, we take an extensive approach into surveying intra-vehicular security, particularly focusing on the inter-ECU communication over the CAN bus (for works that extensively survey the security of V2V and V2I communication over Vehicular Ad Hoc Networks, we refer the readers to \mbox{~\cite{mokhtar2015survey, bernardini2017security, sakiz2017survey, dibaei2020attacks, hasrouny2017vanet, al2012survey, cui2019review, petit2014potential}})}. We start off by reviewing some common attack surfaces that researchers have practically exploited to gain access to the CAN bus. We then look into security measures proposed by researchers over the years and discuss their limitations. Next, we discuss subtle implications to vehicular security when the judgment of a human driver is taken into account. Finally, we share our observation of an efficient, cost-effective and incrementally deployable security solution for the CAN bus and cover its fundamentals.
\\\\

\section{Background on CAN}

\subsection{CAN Frames}
CAN connects nodes along a bus that is broadcast in nature, meaning each message is sent to every node on the bus. Messages do not have a return address. Instead, nodes interpret whether a message is intended for them based on metadata describing what type of data the message holds. A CAN frame can be one of four types: a data frame,  remote frame, an error frame, or an overload frame. A data frame contains data that is to be interpreted or processed by the receiver. A remote frame is used to request transmission of a specific message. An error frame, which starts with a 6-bit error flag, is used to indicate an error has occurred, and an overload frame is used to add a delay between frames \cite{cho2016error}. From a security point of view, data frames (seen below in Figure \ref{CAN-frame}) are the most relevant. 

\begin{figure}[!bth]
	\centering
	\includegraphics[width=\columnwidth]{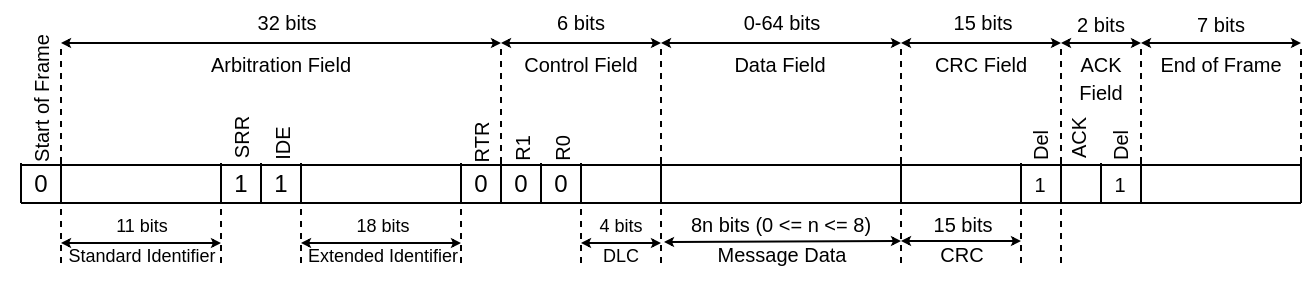}
	\caption{Visual representation of a CAN frame}
	\label{CAN-frame}
\end{figure}

\begin{table}
  \begin{tabular}{ | p{0.2\columnwidth} | p{0.085\columnwidth} | p{0.55\columnwidth} | }
    \hline
    \textbf{Field} & \textbf{Bits} & \textbf{Description} \\ \hline
    Start-of-Frame & 1 & Indicates start of frame \\ \hline
    Identifier 1 & 11 & Standard identifier; contains message priority \\ \hline
    Substitute Remote Request & 1 & Recessive \\ \hline
    Identifier Extension & 1 & Recessive to indicate an extended ID, dominant  otherwise \\ \hline
    Identifier 2 & 18 & Extended identifier; also contains message priority \\ \hline
    Remote Transmission Request & 1 & Dominant indicates standard data frame; recessive indicates RTR \\ \hline
    Reserved Bits & 2 & Reserved bits which are set dominant \\ \hline
    Data Length Code (DLC) & 4 & Indicates number of bytes of data \\ \hline
    Data & 0-64 & Data being transmitted (ranges from 0 to 8 bytes) \\ \hline
    Cyclic Redundancy Check & 15 & Used for error detection \\ \hline
    CRC Delimiter & 1 & Recessive Bit \\ \hline
    Acknowledge\-ment Slot & 1 & Recessive upon transmittal, dominate indicates receipt acknowledgement \\ \hline
    ACK Delimiter & 1 & Recessive Bit \\ \hline
    End-of-Frame & 7 & All recessive bits \\ \hline
  \end{tabular}
  \caption{Description of CAN data frame segments. Here, a \emph{dominant} state is defined as a positive voltage between the CAN bus data wires, while a \emph{recessive} state corresponds to zero voltage between the wires. Unlike a majority of wired transmission protocols, the CAN standard associates a logical zero with a dominant state and a logical one with a recessive state \cite{texas2016intro}.}
    \label{CAN_frame_description}
    \vspace{-3mm}
\end{table}

Table \ref{CAN_frame_description} provides descriptions and bit lengths for each field within a CAN data frame. A data frame can carry up to 8 bytes of data, with the exact data length specified in the Data Length Code (DLC). Additionally, a 1-byte checksum, although not a part of the CAN protocol, is typically contained in the last byte of the data field to ensure integrity \cite{cho2016error}. The identifier field is used to determine if a broadcast frame is useful to a receiving node. The base frame format allows for an 11-bit identifier while the extended frame format allows for a second identifier field containing another 18 bits of identifier data, resulting in a 29-bit identifier. A frame will contain an extended identifier only if the Identifier Extension bit is recessive (the CAN standard associates a logical zero with a dominant state and a logical one with a recessive state).

\subsection{CAN Access Contention}
The identifier field also contains data regarding the priority of a frame, which is used to resolve conflicts between two nodes attempting to transmit data over the CAN simultaneously. Nodes must wait for the CAN bus to become idle before they can transmit data. Once the bus is idle, it is possible that two nodes simultaneously issue frames for transmittal. In this situation, the CAN protocol resolves the issue using an arbitration process based on priority, in which the node that transmitted the frame with the lowest identifier value receives access to the bus. To determine this, the contending nodes send their frames one bit at a time and monitor the output of the bus. The CAN bus acts as a logical \texttt{AND} gate in this situation, \texttt{AND}ing the bits from each frame. If a node observes a dominant (0) bit where it sent a recessive (1) bit, it has lost the arbitration process and the other competing node is given access to the CAN bus. The winner can then broadcast its message over the bus and the loser is put back in  receiving mode where it waits to transmit its message until the bus is detected idle again. Once the arbitration winner has finished transmitting its message, there is a 3-bit buffer before the CAN bus is again open for access -- the node that lost the original arbitration process may now attempt to broadcast its message. This protocol ensures that higher-priority frames are always given first access to the CAN bus \cite{cho2016error, pazul1999controller}, as long as all nodes respect the protocol.

\subsection{CAN Error Handling}
Like most electronic communication, CAN is subject to bit errors. The purpose of the CAN error handling protocol is to ensure proper functionality of the network in the presence of errors. The CAN protocol includes at least five methods for error detection \cite{kvaser-can-error-handling}:

\textbf{Bit Monitoring:} When a node transmits a message over the CAN bus, it also monitors the output of the bus. If the observed output of the bus is different than the message that was transmitted, then a bit error is issued. This, however, does not occur during the arbitration phase that takes place during CAN access contention.

\textbf{Bit Stuffing:} When a node transmits five consecutive bits of the same value, it will add a sixth bit of the opposite value to the sequence. Receiver nodes will then remove this bit upon receipt. If a receiver reads a frame with six consecutive bits of same value, then the receiver issues a stuff error. 

\textbf{Frame Check:} The CAN standard fixes certain parts of each frame to specific values. These areas are the CRC delimiter, the ACK delimiter, End of Frame, and the intermission between frames. If a node on the CAN observes one of these areas to disagree with the standard, it issues a form error.

\textbf{Acknowledgement Check:} When a node receives a message frame, it is expected to send a dominant (0) bit in the Acknowledgement Slot indicating that the message was correctly received, regardless of whether the message was intended for it or not. The transmitter sends a recessive (1) bit in this field when the message is sent. If the transmitter does not observe a dominant bit in the ACK slot, it issues an acknowledgement error.

\textbf{Cyclic Redundancy Check:} Each frame contains a 15-bit CRC. If a receiver observes a different CRC than what it calculates itself, it issues a CRC Error.

\subsection{CAN Error Confinement}
If a node on the bus detects an error, it will immediately transmit an error frame starting with the error flag, except in the case of a CRC error in which the transmission of the error flag is delayed until the completion of the ACK delimiter. Other nodes will then detect the error flag and discard the current broadcast message. Each node contains two counters: a Transmit Error Counter (TEC) and a Receive Error Counter (REC), that are incremented when a transmitter detects a fault in its own message or a listening node detects a fault in an observed message, respectively \cite{kvaser-can-error-handling, copperhill-can-error-flag}. 

There are three states that a node can exist in: Error Active, Error Passive, and Bus Off. A node is in the Error Active state by default in which it will transmit an active error frame (with an active error flag) each time an error is detected and then proceeds to re-transmit the message. When errors are detected by a transmitter, its TEC is incremented by 8. When a receiver detects an error, its REC is incremented by 1. Thus, the TEC is typically incremented at a faster rate than the REC because the transmitter is often the source of the error. With each successfully transmitted message, both error counters are decremented. If either of a node’s error counter exceed 127, that node will enter the Error Passive state in which it will transmit a passive error frame (with a passive error flag) if an error is detected. Unlike the active error flag which consists of 6 dominant bits, the passive error flag consists of 6 recessive bits and therefore, does not interfere with the bus traffic. The error counters of an Error Passive node proceed to increment at normal rates. If either counter value is incremented past 255, the node will enter a Bus Off state in which it will not transmit anything whatsoever over the bus \cite{kvaser-can-error-handling}.

\subsection{Electronic Control Units}
A modern vehicle is comprised of many embedded components, known as the Electronic Control Units (ECUs), controlling various vehicle functionalities. ECUs can control trivial tasks such as opening windows and unlocking doors or more complicated tasks that are vital to vehicle functionality such as anti-lock braking systems and collision prevention systems \cite{nilsson2008vehicle}. 

The ECUs make up the nodes of a vehicle's communication network. ECUs may be present on multiple networks such as Local Interconnect Network (LIN), Media-Oriented Systems Transport (MOST), or the CAN \cite{nilsson2008vehicle}. This paper only focuses on the CAN traffic. Most vehicles contain a high speed bus for critical functions and a low speed bus for non-critical functions. Additionally, there may be bridge nodes connecting the high and low speed bus, making it possible for every ECU on the network to communicate with every other node. Common ECUs present in modern vehicles are compiled in Table \ref{table:ecu-list} \footnote{Pertaining to the fact that vehicle manufacturers have tendency to name a similarly functioning ECU differently from other manufacturers, the list may contain redundancies.} \cite{miller2015remote, miller2013adventures, shukla2016embedded, coverdill1999truck, nissan-aess}. Similarly Table \ref{table:ecu-data-consumption-production} shows typical data produced and consumed by some of the major ECUs on the bus. 

\begin{footnotesize}
\begin{table}
\resizebox{0.49\textwidth}{!} {
\begin{tabular}{| p{\columnwidth} | }
\hline
\vspace{-5mm}
\begin{multicols}{2}
\begin{itemize}[leftmargin=*]
    \item Adaptive Cruise Control Module (ACC)
    \item Air Suspension Control Unit
    \item Air Conditioning Protection Unit
    \item Amplifier (AMP)
    \item Anti-lock Brakes System Module (ABS)
    \item Autonomous Emergency Steering System
    \item Battery Condition Monitor Module (BCMM)
    \item Blind Spot Left Rear Sensor (LBSS)
    \item Blind Spot Right Rear Sensor (RBSS)
    \item Body Control Module (BCM)
    \item Collision Warning Unit
    \item Communication Unit
    \item Data Link Connector (DLC)
    \item Data Loggers Unit
    \item Door Driver Module
    \item Electric Power Steering Module (EPSM/PSM)
    \item Electric Servo Brake Control Unit
    \item Electronic Brake Control Module (EBCM)
    \item Electronic Parking Brake Module (EPBM)
    \item Electronic Shift Module
    \item Engine Control Module (ECM)
    \item External Disc Module (EDM)
    \item Forward Facing Camera Module
    \item Gauge Control Module (GCM)
    \item Headlamp Leveling Module (AHLM) 
    \item Heated Seats Module (HSM)
    \item Heating, Ventilation, Air Conditioning Module (HVAC)
    \item Instrument Panel Cluster/Driver Information Center (IPC/DIC)
    \item Instrumentation Control Unit
    \item Integrated Center Stack Switch Module (ICS)
    \item Lane Keep Assist System (LKAS)
    \item Memory Seat Driver Module (MSM)
    \item Occupant Classification
    \item Occupant Restraint Controller
    \item Park Assist Module (PAM)
    \item Passenger Door Module (PDM)
    \item Power Liftgate Module (PLGM)
    \item Power Management Control (PMC)
    \item Power-train Control Module (PCM)
    \item Pre-Collision System (PCS)
    \item Radio Frequency Hub Module
    \item Radio Module
    \item Remote Control Door Lock Receiver (RCDLR)
    \item Restraint Control Module (RCM)
    \item Security Alarm Unit
    \item Sensing and Diagnostics Module (SDM)
    \item Skid Control
    \item Star CAN C Body Connector (CCB)
    \item Star CAN C IP Connector (CCIP)
    \item Steering Column Control Module (SCCM)
    \item Steering Column Lock Module
    \item Steering Control Module
    \item Telematics Module
    \item Theft Deterrent Module
    \item Tire Pressure Monitoring System (TPMS)
    \item Traction Control Module
    \item Transmission Control Module (TCM)
\end{itemize}
\end{multicols}
\\
\hline
\end{tabular} } % scope of \resizebox ends here
\caption{Common ECUs found in modern vehicles}
\label{table:ecu-list}
\vspace{-5mm}
\end{table}
\end{footnotesize}

{It is becoming increasingly difficult to accommodate the increasing number of ECUs in modern vehicles due to the bandwidth limitation of the CAN bus which is capped at 1 Mbits per second. While an improved in-vehicle communication standard is around the corner, researchers are exploring ways, in the form of CAN extensions, to increase the bandwidth of the CAN bus without requiring significant hardware changes to the bus and the network. An obvious challenge is that such extensions must be backward compatible with the CAN protocol in that existing hardware conforming to the CAN protocol must be compatible with the extension and work without issues. A major advantage of a backward-compatible CAN extension is that dedicated ECUs can utilize higher throughput while other ECUs are still able to communicate normally over the bus. For some of the backward-compatible CAN extensions, we refer the readers to \mbox{~\cite{ziermann2009can+, sheikh2010improving, kang2016high}}.
}

\section{Attack Surfaces and Exploits}
{This section provides a brief overview on vehicular attack surfaces and reviews existing works related to vehicular attacks. First, attacks that are not CAN-focused are briefly discussed followed by attacks that are focused on gaining access to the CAN bus, either via physical access or remotely. This section also reviews common attack surfaces that allow access to the bus. It should be noted that malicious attacks requiring physical access are often impractical as the methods for conducting those attacks are less realistic than for remote attacks.}

\subsection{Non-CAN Attacks}

\subsubsection{Tire Pressure Monitoring System} 
The Tire Pressure Monitoring System (TPMS) consists of sensors inside each tire to monitor the pressure as well as an ECU responsible for communicating with the sensors. The ECU alerts the vehicle operator of an under-inflated tire by sending messages over the CAN to a central vehicle computer, typically a TCM \cite{ishtiaq2010security}. A wired connection from the sensor to the TPMS ECU is not feasible due to the rotating wheel, thus, a wireless communication protocol is used to send the information. Through reverse engineering, the authors of \cite{ishtiaq2010security} were able to learn the proprietary information behind the wireless communication, such as the modulation and encoding schemes and the message formats, and were able to receive and spoof TPMS messages at ranges up to 40m using a cheap antenna and basic low noise amplifier. This attack, however, only gives limited access to the CAN. The publishers of the attack were not able to gain unauthorized access to the TPMS ECU but were instead able to spoof messages to the ECU causing it to alert the driver of low tire pressure despite the pressure being adequate. While this is not necessarily a direct risk to the driver and passenger safety, it poses concerns as to the possibilities of further malicious intent based on a driver's reaction to these messages.

{\footnotesize
\begin{table}
\resizebox{0.49\textwidth}{!} {
\begin{tabular}{ | p{0.10\columnwidth} | p{0.40\columnwidth} | p{0.41\columnwidth} |}
 \hline
 \textbf{ECU} & \textbf{Data consumed from the CAN bus (from)} & \textbf{Data produced to the CAN bus (to)}\\
 \hline
 
ABS %\tablefootnote{Integrated with EBCM.}
&
\textbullet\ Deceleration data (ACC) & 
\textbullet\ Deceleration acknowledgement data (ACC) \\
\hline 

ACCM
& 
\textbullet\ Accelerator pedal position (PCM) \newline
\textbullet\ Vehicle configuration data (BCM) \newline
\textbullet\ Brake pedal applied (PCM) \newline
\textbullet\ Cruise control override (PCM) \newline
\textbullet\ Ignition Status (BCM) \newline
\textbullet\ Steering wheel switch speed control (SCCM) \newline
\textbullet\ Stability control event in progress (ABS) \newline
\textbullet\ Traction Control event in progress \newline
\textbullet\ Vehicle lateral acceleration (RCM) \newline
\textbullet\ Vehicle longitudinal acceleration (RCM) \newline
\textbullet\ Vehicle yaw rate (RCM)
& 
\textbullet\ Adaptive cruise control Brake deceleration request (ABS) \newline
\textbullet\ Adaptive cruise control gap setting (IPC) \newline
\textbullet\ Adaptive cruise control message display (IPC) \newline \\
\hline

PCS \tablefootnote{PCS and LKAS are commonly integrated as a part of the Driving Support ECU in the Toyota Prius.}
& 
\textbullet\ Yaw rate and Acceleration
%(Yaw Rate and Acceleration Sensor)
\newline
\textbullet\ Steering angle (Steering Angle Sensor) \newline
\textbullet\ Vehicle Speed (EBCM) \newline
&
\textbullet\ Pre-collision brake request (EBCM) \newline
\textbullet\ Information and warnings indicating PCS status (Combination Meter Assembly) \newline
\textbullet\ Seat belt operation request (Seat Belt Control ECU) \newline
\textbullet\ Brake assist standby request (EBCM) \\
\hline

LKAS
& 
\textbullet\ Vehicle Speed (EBCM) \newline
\textbullet\ Yaw Rate and Acceleration (Yaw Rate and Acceleration Sensor)
&
\textbullet\ Steering wheel angle (PSM) \newline
\textbullet\ Information and warnings indicating LKA status (Combination Meter Assembly) \\
\hline

PSM
& 
\textbullet\ Engine Speed (ECM) \newline
\textbullet\ Vehicle Speed (EBCM) \newline
\textbullet\ Steering wheel angle (PAM/IPAS) \newline
\textbullet\ Steering wheel angle (LKAS)
& 
\textbullet\ Signal to limit electrical use (HVAC) \newline
\textbullet\ Warning signal indicating malfunctioning or low battery voltage (Combination Meter Assembly) \\
\hline

EBCM \tablefootnote{ABS, Electronic Brake Force Distribution, Brake Assist, Traction Control and Enhanced Vehicle Skid Control are commonly integrated in EBCM (\textit{Skid Control ECU}) in the Toyota Prius. }
& 
\textbullet\ Steering angle (Steering Angle Sensor) \newline
\textbullet\ Accelerator pedal position (ECM) \newline
\textbullet\ Regenerative brake control value (PMC) \newline
\textbullet\ Brake request signals (Driving Support ECU) \newline
\textbullet\ Throttle position (ECM) \newline
\textbullet\ Engine speed (ECM) \newline
\textbullet\ Parking brake switch signal (Main Body ECU)
&
\textbullet\ Warning signal indicating malfunctioning, parking brake on or parking fluid level low (Combination Meter Assembly) \newline
\textbullet\ Regenerative brake signal (PMC) \newline
\textbullet\ Vehicle speed (PSM) \\
\hline

ECM
& 
\textbullet\ Accelerator Pedal Position (PMC) \newline
\textbullet\ Signal of throttle control request \newline
\textbullet\ Engine immobilization signal (Certification ECU)
& 
\textbullet\ Throttle position (EBCM) \newline
\textbullet\ Engine speed (EBCM) \newline 
\textbullet\ Warning signal indicating malfunctioning (Combination Meter Assembly) \\
\hline

Main Body ECU
&
\textbullet\ Remote certification information\tablefootnote{\label{proxy}As a proxy to Remote Engine Starter ECU} (Certification ECU)
&
\textbullet\ Parking brake switch signal (EBCM) \newline
\textbullet\ Start engine signal (Certification ECU) \newline
\textbullet\ Information about each door and the luggage compartment door (Certification ECU) \\
\hline
\end{tabular} } % scope of \resizebox ends here
\caption{Typical data consumed and produced by various ECUs on a CAN bus}
\label{table:ecu-data-consumption-production}
\vspace{-5mm}
\end{table}
}

\subsubsection{KeeLoq Cipher}
KeeLoq is a block cipher with 32-bit blocks that is widely used in remote keyless-entry systems despite its short, 64-bit key size. There are numerous attacks on the cipher employing methods such as the slide, guess and determine, fixed points, and algebraic techniques \cite{courtois2008algebraic, bogdanov2007cryptanalysis, bogdanov2007attacks}. The authors of \cite{indesteege2008practical} were able to craft a more efficient attack based on the slide technique combined with a novel meet-in-the-middle attack. The optimized version of the attack uses $2^{16}$ known plaintexts with a time complexity of $2^{44.5}$ KeeLoq encryptions (528 rounds). The total run-time for the attack is 500 days and can be parallelized across $x$ CPUs for an effective run-time of $500/x$ days.

\subsection{CAN Attacks}

\subsubsection{Media player} Some vehicle media players can be an interesting attack surface to gain access to the CAN bus. The authors of \cite{checkoway2011comprehensive} identified two vulnerabilities in the media player of an experimental vehicle. First, the player has an update capability that automatically recognizes an ISO 9660 formatted CD containing a specifically named file. The system then displays a message on the display and if the user does not respond with the correct input, the media player firmware will be re-flashed with the data on the CD. Second, after reverse-engineering the firmware, the authors of \cite{checkoway2011comprehensive} located a file-reading function that makes strong assumptions about the length of the input. They also discovered that the parser for the Windows Media Audio (WMA) files allows for arbitrary length reads. Together, these two discoveries allow for a buffer overflow attack. The attack is difficult to execute, however. The buffer to overflow is not on the stack, but is instead in the BSS segment with no clear control variables to overwrite. There are also state variables immediately following the segment and arbitrarily overwriting these would crash the system. To execute this attack, the authors developed a debugger that communicates over an unused serial port on the media player. The debugger can then be used to analyze system memory and identify function pointers to overwrite. Finally, they modified a WMA file that exploits the buffer overflow vulnerability and allows CAN packets to be sent across the bus.

\subsubsection{OBD-II} 
The on-board diagnostics (OBD-II) port is used by technicians when servicing a vehicle and, for this reason, it has access to all CAN buses within a vehicle. All vehicles in the U.S. are required to support the \textit{PassThru} standard \cite{checkoway2011comprehensive} which is a Windows based API that provides a software interface to communicate with a vehicle’s internal networks and is typically implemented through having a \textit{PassThru} device that connects directly to a vehicle’s OBD-II port. The authors of \cite{checkoway2011comprehensive} identified two vulnerabilities in the most commonly used \textit{PassThru} device for their undisclosed vehicle. First, anyone on the same network as the device can connect to it with ease. This means that if an attacker can gain access to a dealership or service center’s private network, in whatever way possible, they can connect to the device and communicate directly with the CAN. Second, they discovered it possible to compromise the \textit{PassThru} device itself and potentially install a malware, in which case, it would affect any car the compromised device connects to. 

When the \textit{PassThru} device boots, it broadcasts its IP address and TCP port for receiving client requests. The connection between the device and a client device is unauthenticated so gaining access to the network is the only deterrent. The authors of \cite{checkoway2011comprehensive} discovered an input validation bug within the implementation of an API protocol designed for network configuration that allows an attacker to run shell scripts on the device. An attacker could create a program that connects to a \textit{PassThru} device broadcasting its network information, exploit the input validation bug to execute arbitrary shell commands, and install malicious files designed to send pre-programmed CAN messages to whatever vehicle the \textit{PassThru} device connects to. The attacker could also develop a worm that spreads to other \textit{PassThru} devices on the network, potentially installing malware on hundreds of vehicles at a dealership / service center.

In addition to a \textit{PassThru} device, an ECOM device can also be used to interface with the OBD-II port and read and write to the CAN bus, albeit an adapter may be required for connector compatibility. The authors of \cite{miller2013adventures} customized an ECOM cable to interface with the OBD-II port and gain access to the internal network. They used the accompanying ECOM API to inject both normal and diagnostic CAN packets and control various vehicle functionalities including, but not limited to, the steering, brakes, speedometer readings, lights and horns. Further, they were also able to perform denial of service attacks to limit vehicle functionalities such as the steering.

The OBD-II port also provides an interface to connect after-market dongles that facilitate additional functionalities such as remote control and monitoring. The vulnerabilities in these dongles together with the inherent security limitations of the OBD-II interface provides a means to perform more practical attacks on the CAN bus as highlighted in \cite{wen2020plug}. The authors performed comprehensive analysis of 77 wireless OBD-II dongles and exposed multiple vulnerabilities on each of the dongles. They found that $84.16\%$ of the dongles lacked connection-layer and application-layer authentication allowing for unauthorized access to the CAN bus. Further, $67.53\%$ of the dongles lacked filtering of the undefined CAN messages. Some dongles even allowed over-the-air firmware subversion or extraction. By exploiting the identified vulnerabilities, they were able to perform concrete attacks on a test vehicle, such as disclosing vehicle location, extracting diagnostic data, disabling wireless locking capability and interfering with vehicular controls.

\subsubsection{Bluetooth} 
Most modern vehicles are equipped with Bluetooth functionality for hands-free calling, media, etc. which is typically found in the telematics module. The authors of \cite{checkoway2011comprehensive} were able to reverse engineer the program responsible for handling Bluetooth functionality of a test vehicle. Inside the program, they found an easily exploitable call to \texttt{strcpy}, creating a buffer overflow opportunity for any paired device. Additionally, instead of pairing a new device, an attacker could compromise an already paired device. To demonstrate this, the authors created a Trojan Horse that monitors Bluetooth connections on an Android phone and, if the connecting device is a telematics module, executes the buffer overflow attack and sends a malicious payload to the vehicle. 

Attacks leveraging Bluetooth capabilities are not limited to already paired devices. The authors of \cite{checkoway2011comprehensive} were also able to sniff the Bluetooth MAC address of the vehicle using Bluesniff \cite{spill2007bluesniff} which required a previously paired device be present in the vehicle. As the Bluetooth unit of the test vehicle did not require any user interaction for pairing, they were able to brute-force the PIN and pair a new Bluetooth device. However, the authors note that the rate at which PINs can be tested depends entirely on the response time of the vehicle's Bluetooth stack.

\subsubsection{Wi-Fi}
{The trend towards \textit{smart} devices has found its way into automotive industry. Many vehicles nowadays are equipped with a real-time status monitoring functionality where the real-time updates are provided to the user's smartphone over Wi-Fi. Such features are known to expose additional vulnerabilities. In \mbox{\cite{woo2014practical}}, the authors demonstrate the possibility to exploit the Wi-Fi connectivity of an experimental vehicle to gain access to the internal bus. By installing a malicious diagnostic app on the victim's smartphone and leveraging on the Wi-Fi connectivity between the vehicle and the victim's phone, the authors show that is possible to read the CAN message frames, as well as inject malicious CAN messages to take control over the victim's vehicle.
}

\subsubsection{Telematics Control Module (TCM)}
Long-range wireless access can commonly be associated with the telematics module and its cellular network capabilities. Modern vehicles are equipped with cellular interfaces for phone calls, text messages, and navigation purposes \cite{checkoway2011comprehensive}. Cellular data is routed through a Telematics Call Center (TCC) that is operated by the vehicle manufacturer. Normally, when a call is made to the vehicle, the vehicle will first send a random, three byte challenge packet to the TCC and an authentication timer is started. The TCC then hashes the challenge with an 8-byte pre-shared key to generate a response to the challenge that must be received by the vehicle within 12 seconds of the challenge packet being sent. If the time limit is exceeded or the challenge response is incorrect, the vehicle sends an error packet and ends the attempted connection.

In \cite{checkoway2011comprehensive}, the authors were able to create an artificial TCC through which they were able to communicate with a vehicle, sending arbitrary cellular data packets. Two vulnerabilities within the authentication protocol were discovered that can be compounded with a separate vulnerability within the interface to the TCC. First, the implementation of the random challenge is hardly random--the random challenge generator uses a static seed and is reset whenever the telematics unit starts. Essentially, the random key is the same every time the telematics unit starts, allowing an attacker to easily authenticate with the vehicle. Second, the code tasked with parsing the authentication challenge request responses contains an error that authenticates incorrect responses. For carefully formatted incorrect responses, roughly 1 out of every 256 will be interpreted as correct as a result of this error. This is the case as long as the random key generator is not reinitialized. Further, the interface to the TCC assumes that incoming packets will not exceed 100 bytes. Thus, input lengths are not checked, allowing for a buffer overflow. However, the interface to the TCC only allows for a 21 bytes per second throughput. Given the 12 second limit for a response to the authentication challenge, this vulnerability alone is not sufficient to gain access to the telematics module. Instead the vulnerability in the authentication protocol must be exploited first. After authentication, the timeout window is changed from 12 seconds to 60 seconds, allowing enough time for the buffer overflow attack to be executed.

Similarly, \cite{miller2015remote} documents an entire remote exploit chain to compromise the TCM of a 2014 Jeep Cherokee. The TCM contains a D-Bus message daemon that is used for inter-process communication and, using the appropriate D-Bus service, code can be run using the D-Bus' execute method. The authors state that the easiest step from here is to start an SSH service in order to run commands from a remote terminal, which would allow an attacker to control the radio, HVAC, and other non-CAN related functions that are associated with the TCM.

The telematics module is able to communicate over the CAN using a Renesas V850 chip with a Texas Instruments OMAP-DM3730 SoC (which provides functionalities such as infotainment, navigation and Wi-Fi connectivity) acting as an intermediary between the telematics module's D-Bus service and the V850. Thus, compromising the V850 chip could provide an attacker with CAN access. To do so, the file responsible for updating the V850 needs to be located. Carefully modifying this file in order for the V850 to still accept it as an update file allows the attacker to flash the V850 with the modified firmware, which can be utilized to inject arbitrary CAN messages \cite{miller2015remote}.

The authors of \cite{foster2015fast} discovered vulnerabilities in an aftermarket telematics module that has a standard OBD-II port interface to connect to a vehicle. The TCM includes a mini-USB connector which provides debugging capability and emulates a network adapter. When debugging is enabled, a web server and telnet console listen on ports 80 and 23 respectively. However, both these services did not require any form of authentication. On a more serious security issue, anyone with physical access to the system (and some expertise) could remove the NAND flash chip to read and modify its contents. The authors were able to extract cryptographic keys and certificates from the NAND dump and use it to access the SSH service running on the device. Using the key, they were able to authenticate to the device and read and write files, execute commands and install software to modify functionalities. The authors also found that that the manufacturer of the TCM used the same SSH key on several of their other TCM devices. Moreover, if the IP address is known, the same SSH key can be used to login to the TCM over the web which opens doors for remote exploitation.

\subsection{Attack Outcomes}
Multiple vehicular functionalities can be manipulated once the attacker gains unauthorized access to the network, for instance, by exploiting the data path referenced in Table \ref{table:ecu-data-consumption-production}. Some attacks can have potentially critical impact on the integrity of a vehicle and/or safety of its passengers. (We note that although proof-of-concept attacks were performed on specific vehicle models, there is no reason to believe that similar attacks are not possible across myriad of other vehicle makes and models.) Miller and Valasek enumerate attacks which can take control over the braking system, steering, and throttle \cite{miller2013adventures}. 

In a Toyota Prius, the Pre-Collision System (PCS) can be exploited to directly engage the brakes. Spoofed diagnostic packets designed to test the brakes can be used to engage or disable them entirely in a Ford Focus. Steering can be partially controlled in a Toyota Prius by exploiting the functionalities of the Lane Keep Assist (LKA) or the Intelligent Park Assist (IPAS) systems. Moreover, the Prius is vulnerable to momentary controls over its throttle by replaying packets from the Power Management ECU, the Engine Control Module, or the bridge connecting those two units.
Cho and Shin show how to use the CAN's own error detection and handling to force the shutdown of ``healthy'' ECUs using only several spoofed packets \cite{cho2016error}.
Miller and Valasek demonstrate another denial of service attack on the Power Steering Control Module (PSCM) in a Ford Focus that can entirely disable driver steering assistance, preventing the steering wheel from being turned more than 4 degrees regardless of the amount of force the driver applies \cite{miller2013adventures}.

While there are safeguards that prevent or limit the effects of some of these attacks, judiciously forged sensing packets can fool the safety checks to accept that the various preconditions have been met, e.g. by falsely signaling a low speed reading while engaging the brake system.

\section{Mitigations and Security Measures}

This section provides an overview of different security measures that are proposed to overcome some of the security limitations of CAN. Most of the suggested measures can be broadly categorized into two categories: authentication-based and intrusion detection-based. A summary of the proposed measures can be found in Tables \mbox{\ref{table:auth}} and \mbox{\ref{table:ids}}. {It should be noted that the tables are not meant to highlight the advantages and disadvantages of the proposed measures but rather to summarize the popular mitigation measures and highlight their characteristics.}

\subsection{Security based on authentication}

The authors of \cite{nilsson2008efficient} propose a delayed message authentication based on compound message authentication codes. The proposed scheme compounds every four messages sent from an ECU to another ECU and calculates a MAC for the compounded message. The MAC is then split into chunks of four and sent with the subsequent four messages. The receiver, therefore, requires 4 subsequent messages to verify the authenticity of the 4 preceding messages, delaying authentication. The algorithm used to calculate the MAC is the 3GPP encryption algorithm, KASUMI, used in Cipher Block Chaining Message Authentication Code mode.

CANAuth \cite{van2011canauth} is a backward compatible message authentication protocol for the CAN bus. It utilizes out-of-band transmission through the use of the CAN+ protocol \cite{ziermann2009can+} to perform authentication, which allows for a maximum of 15 bytes for an authentication message. Authentication under CANAuth is a two-step process, starting with key establishment followed by authentication. Key establishment requires that each node on the bus has access to one or more pre-shared keys, one for each group of related messages. The key establishment process is divided into two messages, the first of which is divided into three sections: 8 status bits, a 24-bit counter value, and an 88-bit random number. To begin key establishment, this message is broadcast and all nodes possessing the correct pre-shared key are able to generate a session key using the counter value and the random number using HMAC \cite{bellare1997hmac} with the pre-shared key. The counter value guards against replay attacks. Next, the transmitter broadcasts a second message containing the 8 status bits along with a 112-bit signature comprised of a hash of the session key and the counter value. Now, all receiving nodes are able to validate that the transmitting node knows the session key and is trustworthy. Finally, authentication can take place. An authentication message again contains the 8 status bits, a new 32-bit counter value, and an 80-bit signature comprised of a hash of the session key and the new counter value.

LiBrA-CAN \cite{groza2012libra} is a lightweight broadcast authentication protocol designed to address the shortcomings of CANAuth, for instance the impracticality of storing a key for each CAN ID, and uses a progressive authentication mechanism based on \textit{key splitting} and \textit{MAC mixing} paradigms. \textit{MAC mixing} allows for the integration of multiple authentication codes while \textit{key splitting} increases the entropy of each mixed MAC. The Linearly Mixed MACs increase the security as one wrong MAC corrupts all other MACs and thus the verification of the mixed MAC fails on each of the associated keys. The scheme uses a centralized authentication setup consisting of a master node and slave nodes connected to the CAN bus. All slave nodes register to the master node as a part of the key sharing process and the master node distributes the keys. Multiple tags, generated by a tag generation algorithm, are concatenated to build the Mixed MAC.
When the master receives a data frame containing a message from a slave, it checks if the integrated counter  is up to date and queues the message for authentication. Then, when it receives an authentication frame containing a tag from the slave, it takes the matching packet off the queue and authenticates it. If the authentication is successful, it then authenticates the tag to other nodes.

In \cite{groll2009secure}, the authors formulate a security mechanism based on Trusted Communication Groups and a Key Distribution Center (KDC). The KDC generates and transmits group keys to ECUs in each communication group. The protocol uses asymmetric key cryptography for the key distribution phase and symmetric key cryptography to encrypt subsequent CAN messages. Each ECU stores its private key in a tamper-proof memory while the public key of the KDC is made available to all the ECUs. The membership of the group is defined using the ECU's Access Control List (ACL) which is cryptographically signed by the vehicle manufacturer to provide for its integrity. Each ECU can only transmit messages to other ECUs in its group and since the traffic is encrypted, the protocol achieves both authenticity and confidentiality. 

Likewise, the authors of \cite{dariz2017trade} propose an authentication scheme that computes an integrity tag for a given message by hashing the message with an authentication key. The integrity tag is then concatenated with the message and encrypted with an encryption key (\textit{MAC-then-encrypt}). The authors also compare the safety and security properties of other schemes for message protection, namely $ENC+CRC$, $plain+MAC$ and $plain+CRC$, against the proposed $MAC+ENC$ scheme. They conclude that although encryption has stronger security properties, it directly influences the probability of residual error and therefore may interfere with safety.

Another authentication framework, VeCure \cite{wang2014vecure}, is based on a concept of a \textit{trust group}. Each ECU is assigned a trust level based on how easy of a target the ECU is for an attacker, and then ECUs are grouped based on these trust levels. While the high-trust group nodes are able to compute authentication codes and therefore share a secret symmetric key, the low-trust groups are not provisioned with this capability. In the initialization phase, each ECU is assigned a unique 1-byte node ID which is used in the generation and verification of authentication codes. The node IDs along with the symmetric key are stored in the flash memory of each ECU in the high-trust group.
Each data message from the high-trust group is followed by an authentication message that embeds the authentication information. Two bytes of the authentication message are used for the message counter. The message counter together with a session number is used to protect against replay attacks. The session number is initialized for each driving session and stored on the ECU's flash memory. One-byte node ID, 4-byte message authentication code and 1-byte authentication marker make up the remaining bytes of the authentication message. The computation of the message authentication code is carried out in two phases: a heavyweight offline computation and a lightweight online computation. The offline computation, carried out in advance, is a hash of the node ID, the session number, the overflow counter, the message counter and the symmetric key, but not the data. The data is a parameter to the online computation along with the hash to compute the final MAC.

In a more recent work, Kurachi et al. suggest a centralized authentication system for the CAN bus: CaCAN \cite{kurachi2014cacan}. CaCAN introduces the concept of a monitor node, a node tasked with authenticating other nodes on the bus. The monitor node and each ECU share cryptographic keys which are used for computing the message authentication code. The authorization keys are stored in an anti-tamper memory of the monitor node. This centralized authentication system requires a hardware modification to the CAN bus as the monitor node requires a special CAN controller, HMAC-CAN. Every data frame sent on the bus has a MAC that is checked by the HMAC-CAN controller. If the HMAC-CAN controller detects an unauthorized message, it overwrites the unauthorized frame with an error frame in real time, destroying the unauthorized message and eliminating its unwanted effects. {Similar authentication schemes are also discussed in \mbox{\cite{ueda2015security}} and \mbox{\cite{wang2017hardware}}}.

LeiA \cite{radu2016leia} is another fully backwards compatible authentication protocol for the CAN bus. In this protocol, each participant stores a tuple consisting of the CAN ID, a 128-bit long-term symmetric key used to derive the session key, a 56-bit epoch that contributes in the generation of the session key, a 128-bit session key used to generate the MAC and a 16-bit counter value embedded in the MAC and sent with the messages. The sender and the receiver first generate the session keys using the long-term key and the epoch. The epoch is incremented each time before a session key is generated and the counter is set to zero after the generation of a session key. Before sending an authenticated message, the sender updates the counter and if required, the epoch. The sender then computes the MAC with the session key, the counter and the data and then transmits the counter, the data and MAC. After receiving these values, the receiver verifies the MAC. The protocol also allows for \textit{resynchronization} if the MAC cannot be verified.

LCAP \cite{hazem2012lcap} is a lightweight authentication protocol for CAN that relies on the use of a 2-byte magic number. The number is computed using the hash function used in the TESLA \cite{perrig2000efficient} protocol and appended to each message. To compute the magic number, the sender picks a random number and repeatedly applies a transformation function. The initial magic number of each message is broadcast to all receivers. As using the same magic number to authenticate all the messages from a sender leaves a big security hole, the protocol, instead, orders messages such that the first message can be verified by applying the hash function once, the second message by applying the hash function twice, and so forth. The protocol also defines two modes of operation: \textit{Extended Mode}, in which the \textit{Extended Identifier} field of the CAN message is used to send the magic number, and \textit{Standard Mode}, in which the magic number is sent in the payload--thus, consuming 2 bytes of the payload--and the whole payload is encrypted using a symmetric key.

vatiCAN \cite{nurnberger2016vatican} is another authentication mechanism for CAN which uses a separate CAN message for authentication purposes. An authentication message with a different sender ID follows a critical message to be authenticated. However, unlike other similar mechanisms, only selected messages are authenticated, thereby significantly reducing the overhead of authentication. Also, only vatiCAN-aware recipients authenticate the critical message. As with other similar mechanisms, the execution of the corresponding command is deferred until the reception of the authentication message for the corresponding critical message. Messages that fail authentication are discarded.

{LEAP \mbox{\cite{lu2019leap}} attempts to overcome the computational costs of the MAC-based authentication by using a stream cipher, \textit{RC4}, to encrypt and authenticate CAN messages. It uses a dedicated ECU to store the long-term symmetric keys used for generating the session keys. The sets of session keys are only common to the ECUs in the same communication group and updated periodically to impede brute-force attacks. The RC4 algorithm is used to generate a key-stream using the session key. For the purpose of authentication, the sending ECU's ID (11-bit) is encrypted with a part of the key-stream. The plaintext message, consisting of the actual data and the encrypted id, is encrypted using another part of the key-stream and the resulting ciphertext is sent in the data field of a regular CAN message. At the receiving end, the receiver ECU generates the same key-stream using the shared session key. The ciphertext is then decrypted and the id of the received CAN message is compared against the decrypted id from the ciphertext. If they are the same, the authentication is successful; else, the authentication fails, and the message is discarded.}

\subsection{Security based on intrusion detection}\label{intrusion_id}
{Many types of intrusion detection systems have been recently proposed in the literature, for various types of applications (see, for instance, \mbox{~\cite{radoglou2019securing, baykara2018novel, butun2013survey, baykara2017novel, baykara2015survey}}). However, the limitations of the CAN protocol and the CAN-conforming hardware makes it difficult to readily adapt robust intrusion detection measures from other domains to vehicular security. Several researchers have nonetheless attempted to develop compelling intrusion detection systems for the CAN bus. In this section, we focus on some of the main intrusion detection systems designed to detect anomalies in the CAN traffic and secure the CAN bus from successful attacks.
}

In \cite{hoppe2011security}, Hoppe et al. discuss three different tests for intrusion detection: increased message frequency, obvious misuse of message IDs, and low-level communication characteristics. Message frequency techniques are based on the observation that many attacks involve repeatedly injecting packets to the CAN bus which results in higher than normal frequency of the corresponding packets. This same approach is further developed in \cite{gmiden2016intrusion}. On the other hand, misuse of message IDs refers to the fact that attackers often compromise a node (typically, an ECU) to inject packets that look like packets from some other nodes. As CAN is a broadcasting protocol, the sending node also receives the message, but it is not expected to evaluate it. As such, a simple source ID functionality can be added to each node to check for whether the node actually generated the current message bearing an ID exclusively used by that node. Lastly, detection patterns that involve low-level communication characteristics are based on observing electrical signals in the physical layer. When ECUs generate CAN messages, the CAN controller generates electrical signals on the bus to broadcast the message. The signals generated may act as a fingerprint of the source ECU. {This insight has been realized in \mbox{\cite{murvay2014source} and \cite{choi2018identifying}} where the unique electrical characteristics of the ECUs are analyzed in the physical layer to identify the source of a current message (for the purpose of authentication) \mbox{\cite{murvay2014source}} and to detect maliciously acting ECUs (for the purpose of intrusion detection) \mbox{\cite{choi2018identifying}}}.

Some other similar intrusion detection methods that use the frequency of CAN messages as a detection pattern include the works in \cite{song2016intrusion} and \cite{taylor2015frequency}. The hybrid IDS proposed by the authors of \cite{song2016intrusion} uses attack signatures and anomalies in CAN traffic frequency to detect possible attacks. Primarily, the IDS keeps a score of anomalies in message frequency and whenever the score hits a preset threshold, the event is identified as an attack. Similarly, the flow-based anomaly detection scheme discussed in \cite{taylor2015frequency} uses a sliding window approach that computes the flow of the CAN packets in the preset window and compares against a historical reference to detect anomalies. The authors also explore the effectiveness of the approach over a range of packet injection frequencies to determine its practical limitations, and point out that, while the timing information can be reliably used to detect anomalous packets, the Hamming distance between successive packet data fields is not a good anomaly indicator.

More concrete IDS models based on message frequency have been recently developed. One such model with a very high true positive to false positive ratio is discussed in \cite{moore2017modeling}. It observes the CAN data for a few seconds and records the timing information of CAN messages. It then uses this timing information against future observations to detect anomalies and potential attacks. The model also accounts for the possibility that even during normal events, CAN messages may be lost because of collisions. Hence, the observed timing information may be different from the recorded timing information. To reduce the number of false positives, the model requires three consecutive anomalies before an alert is issued. 

An anomaly detection approach that uses entropy to specify the normal behavior of the vehicular system and change in entropy as a potential attack is discussed in \cite{muter2011entropy}. The scheme is based on the fact that automotive networks are restrictive in nature: each packet and its potential content is pre-specified, the ID of a CAN message is correlated with the semantics of the payload, and the frequencies of many messages are well defined. Put simply, the vehicular system contains a low entropy, while any attacks injecting new packets or manipulating the payloads of regular packets increase the entropy. 

Machine learning models have also been used to detect anomalies in the CAN bus. In \cite{kang2016intrusion}, the authors propose using a deep neural network (DNN) to detect attack packets. The DNN takes the data fields of CAN packets as inputs and outputs a binary label that identifies the packet as normal or malicious. {A similar model can be found in \mbox{\cite{amato2021can}} which, in addition to identifying malicious packets, classifies the injected packets into four attack types: denial-of-service attack, fuzzy attack, drive gear spoofing and RPM gauge spoofing}. Similarly, the authors of \cite{markovitz2017field} use a machine learning model to detect deviations in CAN traffic. The proposed system uses a classifier to identify the field types of the CAN messages. Once the field types are identified, the messages are fit into a model similar to the Ternary Content-Addressable Memory (TCAM) model. A set of TCAMs is created for each message ID and all messages that meet the properties of that message ID's fields are grouped in the same set. Any messages that do not fit into a set are considered anomalous. Likewise, \cite{marchetti2017anomaly} proposes an anomaly detection algorithm based on the analysis of CAN message sequences. 
The algorithm proceeds with its training phase by building a reference model based on the identification of recurring patterns in CAN message IDs during its normal operation. The observed transition between consecutive message IDs is captured in a data structure called the \textit{transition matrix} -- in essence a simplified Markov model. In the detection phase, the current sequence of message IDs is validated against the transition matrix. If any message ID transition does not appear in the transition matrix, then the validation fails.

{An intrusion detection architecture relying on packet sequences is discussed in \mbox{\cite{taylor2016anomaly}}. The mechanism uses a Long Short-Term Memory (LSTM) Recurrent Neural Network (RNN) to predict the next packet data values from each sender in the bus. Any inconsistency in the observed data versus the predicted values indicate an anomaly in the data traffic. The model has the benefit that it does not involve decoding the CAN messages, and therefore does not require knowledge of the message semantics.}

{Recently, an intrusion detection system using \textit{Transfer Learning} was proposed in \mbox{\cite{tariq2020cantransfer}}. It uses a two-step learning process: first, a Convolutional LSTM (ConvLSTM) model is trained using normal and known attack data. A dataset, initially consisting of four features of CAN packets: timing information, ID, DLC and the data, is pre-processed and transformed to a time series, which is subsequently transformed into two-dimensional spatial data for the training process. Next, one-shot transfer learning is used to retrain the model to detect new attacks. The advantage of using Transfer Learning is that the model can be retrained quickly and new attacks can be detected using only a small number of new data points. 
}

\begin{table}[ht]
%\resizebox{0.49\textwidth}{!} {
\begin{tabular}{c c c c c c c c}
Paper & \rot{90}{Relies on CAN+} & \rot{90}{Central Authentication} & \rot{90}{Specialized Hardware} & \rot{90}{MAC Length} & \rot{90}{Asymmetric Key} & \rot{90}{Real time Authentication} & \rot{90}{Trust Group Based} \\
\hline
 \cite{nilsson2008efficient}  & \no & \no & \no & 64 bits & \no & \no & \no \\ \hline
 \cite{van2011canauth}  & \yes & \no & \no & 15 bytes & \no & \yes & \no \\ \hline
 \cite{groza2012libra}  & \sorta & \yes & \yes & 64 bits (max) & \no & \sorta & \no \\ \hline
 \cite{groll2009secure}  & \no & \no & \yes & Undefined & \yes & \yes & \yes \\ \hline
 \cite{dariz2017trade}  & \no & \no & \no & Undefined & \no & \yes & \no \\ \hline
 \cite{wang2014vecure}  & \no & \no & \no & 32 bits & \no & \sorta & \yes \\ \hline
 \cite{kurachi2014cacan}  & \no & \yes & \yes & 8 bits & \no & \yes & \no \\ \hline
 \cite{radu2016leia}  & \no & \no & \no & 64 bits & \no & \sorta & \no \\ \hline
 \cite{hazem2012lcap}  & \no & \no & \no & \na & \no & \yes & \no \\ \hline
 \cite{nurnberger2016vatican} & \no & \no & \no & 64 bits & \no & \sorta & \sorta \\ \hline
 {\mbox{\cite{wang2017hardware}}} & \no & \yes & \yes & 3 bytes & \no & \yes & \no \\ \hline
 {\mbox{\cite{lu2019leap}}} & \no & \no & \yes & \na & \no & \yes & \no \\ \hline
\end{tabular} % } % Scope of resizebox ends here
\caption{Summary of proposed backward-compatible authentication mechanisms for the CAN bus \\
\yes \ Yes \hspace{1cm} \no \ No \hspace{1cm} \na \ N/A \hspace{1cm} \sorta \ Partial}
\vspace{2mm}
\label{table:auth}
\end{table}

\begin{table}[ht]
%\resizebox{0.49\textwidth}{!} {
\begin{tabular}{c c c c c c c c c c c}
Paper & \rot{90}{Signature-based} & \rot{90}{Anomaly-based} & \rot{90}{ML model} & \rot{90}{Examines Packet Frequency} & \rot{90}{Examines Packet Payload} & \rot{90}{Examines Packets Sequence} \\
\hline
 \cite{hoppe2011security} & \no & \yes & \na & \yes & \no & \no  \\ \hline
 \cite{gmiden2016intrusion} & \no & \yes & \na  & \yes & \no & \no  \\ \hline
 \cite{song2016intrusion} & \yes & \yes & \na & \yes & \no & \no  \\ \hline
 \cite{taylor2015frequency} & \no & \yes & \na  & \yes & \no & \no  \\ \hline
 \cite{moore2017modeling} & \no & \yes & \na  & \yes & \no & \no  \\ \hline
 \cite{muter2011entropy} & \no & \yes & \na  & \sorta & \sorta & \no \\ \hline
 \cite{kang2016intrusion} & \no & \yes & Deep Neural Network  & \no & \yes & \no \\ \hline
 \cite{markovitz2017field} & \no & \yes & TCAM  & \no & \yes & \no \\ \hline
 \cite{marchetti2017anomaly} & \no & \yes & Transition Matrix  & \no & \no & \yes \\ \hline
 \multirow{2}{*}{{\mbox{\cite{taylor2016anomaly}}}} & \multirow{2}{*}{\no} & \multirow{2}{*}{\yes} & Long Short-term Memory & \multirow{2}{*}{\no} & \multirow{2}{*}{\yes} & \multirow{2}{*}{\yes} \\
    & & & Recurrent Neural Network & & \\ \hline
 {\mbox{\cite{amato2021can}}} & \no & \yes & Deep Neural Network & \no & \yes & \no \\ \hline
 {\mbox{\cite{tariq2020cantransfer}}} & \no & \yes & Convolutional LSTM Network & \yes & \yes & \no \\ \hline
\end{tabular} %} % Scope of resizebox ends here
\caption{Summary of proposed backward-compatible intrusion detection systems for the CAN bus \\
\yes \ Yes \hspace{0.1cm} \no \ No \hspace{0.1cm} \na \ N/A \hspace{0.1cm} \sorta \ As a part of other statistics}
\label{table:ids}
\end{table}

\subsection{Limitations of existing security measures}
An effective solution to the CAN security issues needs to be cost-effective, meet the real-time communication needs of the vehicle and be scalable, in that an increase in the number of ECUs should not significantly hinder the performance, impact security or increase cost. It must also be compatible across vehicles from different manufacturers. Moreover, while an improved standard is around the corner, the solution must be backward-compatible with the current CAN specification.

Unfortunately, the numerous suggested solutions for securing the in-vehicle network suffer from various limitations, and an incomplete understanding of how well they would be able to meet the performance, security and cost needs of a vehicle. For instance, most of the authentication measures proposed in the existing literature require a second authentication packet that follows a data packet. Clearly, the authentication is delayed until the reception of this packet. This engenders latency in communication and impacts the real-time communication needs of the vehicle. Further, additional CAN messages for authentication increase the residual error rate \cite{dariz2017trade, iso2008earth}. Moreover, some authentication mechanisms need specialized central gateways which results in an increased cost of production. Authentication-based countermeasures are also mostly ineffective against DoS attacks as they do not prevent an attacker from flooding the bus with a pool of CAN messages. This underlying shortcoming can be attributed to the fact that these countermeasures do not prevent an attacker from injecting messages to the bus, i.e. they do not prevent the production of CAN messages, rather, by design, they only prevent consumption and usage of maliciously injected messages. Similarly, anomaly detection based on frequency of CAN messages suffers from the problem that non-periodic packet types are not handled properly. Further, data fields of packets are not examined, only the timing, which makes the solution much less robust. {In addition, most of the proposed intrusion detection systems have either missing accuracy evaluations or questionable accuracy as the accuracy of the model is often evaluated on a small number of mostly synthetic datasets. It is also not clear if the anomaly detection patterns used in the models are effective across different vehicle makes as CAN packets significantly differ in payload and packet-sequence across different manufacturers. It has also been demonstrated that a carefully crafted DoS attack can be mounted on the CAN bus even in the presence of an intrusion detection system that analyzes CAN messages to detect potential attacks \mbox{\cite{palanca2017stealth}}. More importantly, researchers have recently devised attacks that generate error patterns \mbox{\cite{kulandaivel2021cannon}} that are indistinguishable from normal CAN errors and therefore, can elude all contemporary intrusion detection systems.}

\begin{figure}[tb]
%\vspace{-2mm}
	\centering
	\includegraphics[width=0.9\columnwidth]{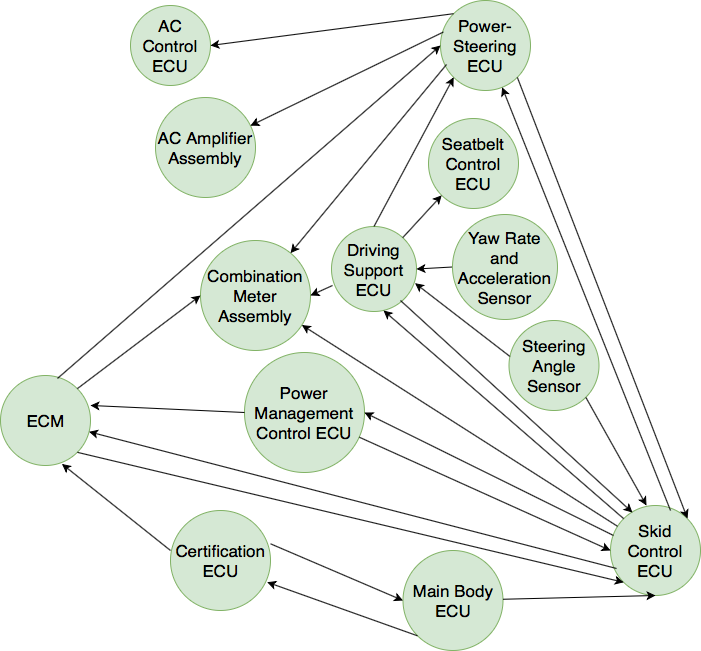}
	\caption{Adjacency graph reflecting the data flow (directed edges) between ECUs (nodes)} %\vspace{-4mm}
	\label{ecu-adjacency-graph}
	\vspace{-2mm}
\end{figure}

\section{Human-in-the-Loop} \label{human-in-the-loop}
{Most of the analyses performed in the domain of vehicular security consider the communication network, CAN, made up solely of the common vehicle ECUs. In consequence, prior works have largely overlooked a critical node in the CAN communication network: the driver. The human driver becomes relevant when discussing inter-ECU communication as the driver acts as a virtual node in the network providing additional communication paths between ECUs. As the automotive system is essentially a control system, the driver, being a central entity in the control loop, has a direct control over most vehicle functionalities. It may be possible to exploit this control to carry out attacks that are otherwise not feasible. This is especially concerning as the attacks can be executed even in the presence of some security measures.}

To better elucidate the subtle implications of a human driver's involvement in the vehicular security, we consider a setting where an attacker has an unrestricted access (for instance, by exploiting a vulnerability) to the Radio Module of a car. Assuming the volume packets from the Radio Module are sent over the CAN bus, the attacker can craft and inject false volume packets to suddenly turn up the volume. Such an action could distract a driver and cause serious accidents, especially during a simultaneous stressful event (say, hard braking). Notice that this attack can be successfully executed even in the presence of an Intrusion Detection System (especially, if the detection is based on packet frequency and packet type) as the injected packets in this setting are likely to pass all legitimacy tests.

\begin{figure}[tb]
%\vspace{-2mm}
	\centering
	\includegraphics[width=0.5\columnwidth]{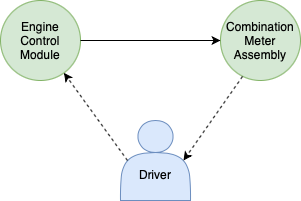}
	\caption{Virtual path from the Combination Meter Assembly to the Engine Control Module} %\vspace{-4mm}
	\label{ecm_cma_driver}
	\vspace{-2mm}
\end{figure}

The existence of the human node also creates novel data paths. To illustrate this, we refer the reader to the \textit{ECU Adjacency Graph}\footnote{Based on the information obtained from the service manuals and data-sheets of various Toyota vehicles} shown in Figure \ref{ecu-adjacency-graph}. For simplicity, we omit information about the actual data that flow between ECUs. In the figure, we observe no data path from the Combination Meter Assembly (CMA) to the  Engine Control Module (ECM). Therefore, it is not feasible for an attacker to influence the ECM by compromising the CMA. However, with the introduction of a human driver in the control loop, it becomes possible for the attacker to leverage on the driver's judgment to create a virtual path from the CMA to the ECM as shown in Figure \ref{ecm_cma_driver}. For instance, by displaying a false speedometer reading on the display, the attacker could motivate the human driver to accelerate or decelerate the car. Note that the attacker does not have to inject false speed packets (e.g., by mimicking the Adaptive Cruise Control), rather she only has to display a false reading on the dashboard. Although the CMA is an unlikely attack vector, we stress that the concept holds true in general. In essence, with human involvement, the possibilities are unlimited.

The consideration of the human in the network exposes additional vulnerabilities of the overall system. The notion of human-in-the-loop therefore becomes pertinent when devising future-proof security solutions for CAN. 

\section{The ``inverted firewall'' solution}

Despite the large pool of proposed security solutions for CAN, a solution that is also resistant to DoS attacks, and yet simple and cost-effective, appears to be missing from the pool. A majority of known CAN attacks involve injecting packets either locally or remotely; therefore, it is only logical to look for a solution that involves filtering packets from unknown or compromised sources. However, CAN packets do not carry source information (see Section~\ref{intrusion_id}) and consequently source-based filtering is not possible unless additional metadata is incorporated into the payload itself, making firewall-like solutions difficult without significant changes to the CAN protocol.

Notably, the majority of packet injection attacks on CAN involve compromised ECUs mimicking some other ECUs in the network. It is rarely the case that ECUs that are directly responsible for an action such as controlling vehicular speed are compromised. Rather, an attacker compromises some other ECU and starts injecting packets to mimic one or more ECUs that are responsible for the action. For instance, in an adaptive cruise control mode, the Engine Control Module (ECM) and the Adaptive Cruise Control System (ACCS) are two ECUs responsible for controlling vehicular speed. However, in none of the documented attacks, either of these ECUs is compromised. Instead, some other ECU, such as the TCM, is compromised and packets are injected from it to mimic the ACCS. By enforcing a rule that the ECUs can only produce packets that meet some predetermined specifications, such mimicking behaviour can easily be counteracted. This observation immediately implies the need for a system that filters packets at the source ECU rather than the destination -- a kind of inverted firewall -- which for lack of a better term we call \emph{icewall}.

\begin{figure}%[htb]
	\centering
	\includegraphics[width=1\columnwidth]{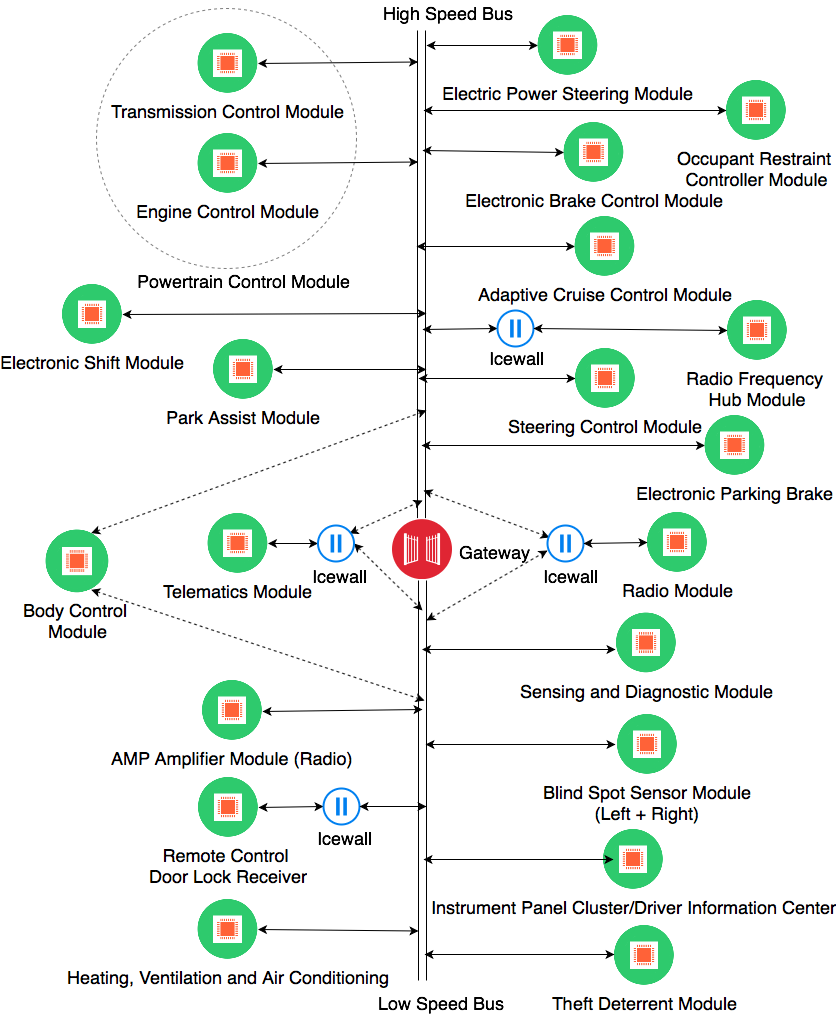}
	\caption{An example icewall installation on a CAN bus}
	\label{icewall-installation}
\vspace{-5mm}
\end{figure}

An icewall can be installed between an ECU and the CAN bus, filtering all outgoing ECU packets before they are transmitted. An example installation of multiple icewalls, where each icewall monitors an ECU that potentially exposes a remote interface, is shown in Figure \ref{icewall-installation}. At its heart, an icewall monitors the corresponding ECU and ensures that all packets originating from the monitored ECU comply with the preset rules regarding the content of the packet. All outgoing packets that do not meet its specifications are blocked and never make it to the bus, thereby eliminating any potential mimicking behavior. It should be noted that, by design, an icewall is supposed to let all incoming packets through.

An icewall is inherently resistant to many DoS attacks. This is attributed to the fact that an icewall prevents compromised low priority ECUs from injecting high priority CAN messages and further, can be configured to limit the frequency of outgoing messages. With additional filter rules that disallow continuous injection of active error frames once the threshold (as defined in the standard) is hit, icewalls can be deployed as a viable solution to defend against the majority of DoS attacks on the bus.

\subsection{Attacker Model}
As an icewall is based on packet filtering, it can be effective against a class of false packet injection attacks. Considering a typical attack constraint, it would be reasonable to assume that an attacker is unable to compromise ECUs directly responsible for safety-critical actions (for instance, the ECM or the ACCS responsible for controlling the speed of a vehicle) as, to the best of our knowledge, all practical documented attacks involve remotely compromising a secondary ECU (or a set of ECUs) and injecting packets to mimic behavior of other ECUs \cite{miller2015remote, nie2017free}.

It would also be reasonable to assume that the attacker has remote wireless access to the vehicle (and thus can interact with, and possibly compromise, any ECUs with wireless communication capabilities), and that the only direct physical access is limited to the OBD-II port. In this case, an icewall can be an effective defensive measure, primarily because it can be installed only where needed rather than at every ECU like a firewall. For instance, if the attacker can only access a vehicle remotely, icewalls can be installed for all those ECUs that expose remote interfaces, as shown in Figure \ref{icewall-installation}. Similarly, if the attacker can also inject packets through OBD-II port, the OBD-II port itself can be secured with an icewall.

A non-hardware-controlling adversary is also a reasonable assumption in this case, since any attacker with direct access to the vehicle's components would be able to physically remove an icewall (or a firewall), or replace the entire ECU-icewall combination with a malicious device. Such an attacker may further reverse-engineer any ECU, or directly recover any secret keying material from the ECU's non-volatile memory or key generation component, thus rendering most of the proposed defense mechanisms futile. Moreover, any potential defense mechanism that could deal with such an attacker would most likely be undesirable, as it would interfere with vehicle manufacturing, servicing and after-market customization.

\subsection{Icewall Configuration}
An icewall device can be installed as an OEM device or as an aftermarket accessory. Further, an icewall can be manually or automatically configured; the manual configuration of the icewall requires preprogramming filter rules into an icewall device where modifying filter rules can only be done by the vehicle manufacturer or an authorized service personnel. Automatic configuration of an icewall can be achieved by enabling learning abilities into the icewall where it examines the first few incoming packets and sets a filtering rule that permits only those types of packets to enter the bus. An automatically configured icewall, in essence, is an Intrusion Detection System with machine learning abilities. Also, for automatic configuration purposes, an icewall device must come equipped with a reset button that allows it to flush its current filter rules and relearn new rules.

The automatic configuration of the icewall does not come without its own problems, however. If an icewall is installed to monitor an ECU that sends multiple packet types, the installed icewall, in its learning phase, may be unable to learn all types of packets that should be allowed to enter the bus. All other packet types, even the legitimate ones, will therefore be blocked. In this case, for those ECUs, manual configuration of the icewall is desirable. Automatic configuration of icewall does have its benefits, however. For one, icewall devices can be manufactured as universal plug-n-play devices. This clearly reduces the cost of a device as well as the setup effort.

\subsection{Limitations and Enhancements}
One potential limitation of an icewall is that it cannot prevent malicious packets that meet the preset rules from entering the bus. As discussed before, in a typical setting, this does not carry significant safety ramifications. However, when human-in-the-loop is considered, this limitation can have noteworthy implications as discussed in Section \ref{human-in-the-loop}.

Various enhancements to an icewall are possible that not only limit potential damages in the human-in-the-loop setting (Section~\ref{human-in-the-loop}), but also make an icewall more robust. One potential enhancement is to configure an icewall such that it not only examines the type of an outgoing packet but also its payload. This is particularly handy to detect abnormal readings in the packet data and block the packet. For instance, considering the attack scenario in which the attacker displays false speedometer reading on the display to motivate the driver to react in an unsafe manner, the attack can be neutralized by detecting abnormal speedometer readings. For regular cars, it is improbable that a car abruptly accelerates in a fraction of a second. Malicious speedometer display can therefore be potentially characterized by abnormal acceleration readings.  By configuring icewall such that it blocks all packets with abnormal readings, it is therefore possible to limit the consequences of imperfect human judgment.

\section{Conclusion}
The security limitations of the CAN bus communication protocol can be attributed to the fact that the CAN standard was primarily developed to meet real-time communication needs and lacks security controls. The tight timing and packet size limitations of the protocol hinder the development of a simple, cost-effective and efficient security solution. In this work, we reviewed security threats and countermeasures for the CAN bus communication protocol. Our review of the existing literature lead us to
introducing the notion of \textit{human-in-the-loop}, and we discussed subtle implications to security not previously addressed. We also discussed the limitations of existing measures and shared our insights regarding a cost-effective, secure and incrementally deployable solution: an inverted firewall. Referred to in this paper as the \textit{icewall}, the inverted firewall can be effective against a major class of packet injection and denial of service attacks.

\bibliographystyle{IEEEtran}
\bibliography{references}

\begin{IEEEbiography}[{\includegraphics[width=1in,height=1.25in,clip,keepaspectratio]{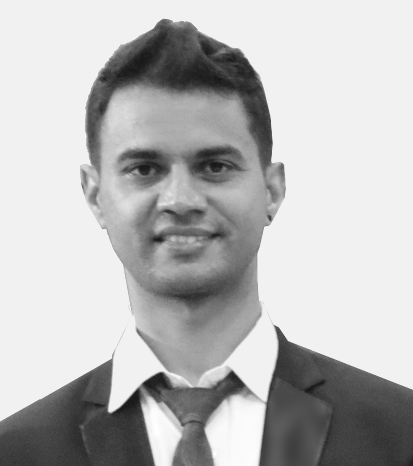}}]{Chandra Sharma} received his BS degree in Computer Engineering from Kathmandu Engineering College, Nepal, in 2015. He joined Kansas State University, KS, in 2017 to pursue the PhD degree in Computer Science.

Starting 2018, he has been working as a Research Assistant at PITS Lab at Kansas State University. His primary research interests include information theory and system and software security. His current work focuses on privacy concerns of disclosing personal information on the internet and optimizing the trade-off between information privacy and utility.

\end{IEEEbiography}

\begin{IEEEbiography}[{\includegraphics[width=1in,height=1.25in,clip,keepaspectratio]{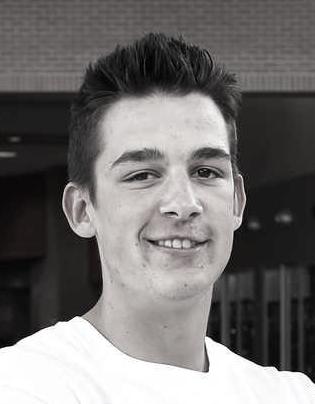}}]{Samuel Moylan} graduated from the department of Computer Science at Kansas State University in 2019. While at Kansas State University, his research interests included vehicular security and cyber security.
\end{IEEEbiography}

\begin{IEEEbiography}[{\includegraphics[width=1in,height=1.25in,clip,keepaspectratio]{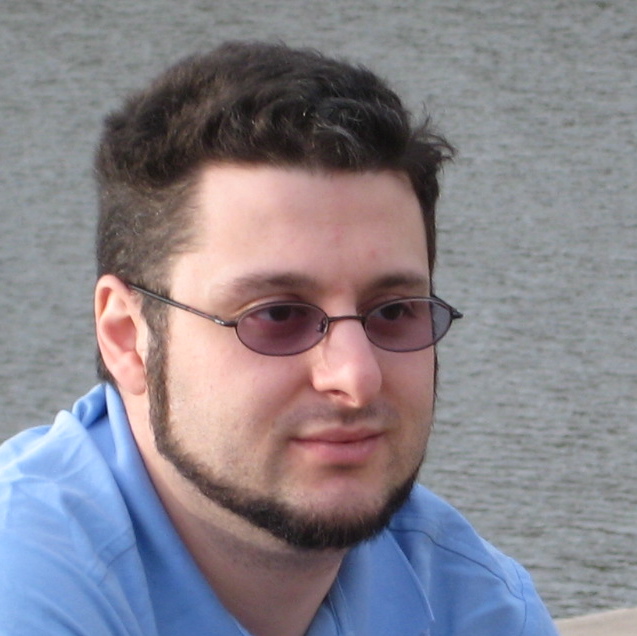}}]{Eugene Y Vasserman} is an Associate Professor in the Department of Computer Science at Kansas State University, specializing in the security of distributed systems. His current research is chiefly in the area of security for medical cyber-physical systems, security usability, and user education, with past work spanning the gamut from medical system authorization with integrated break-glass capabilities (IoMT), to secure hyper-local routing and social networking, to privacy and censorship resistance on a global scale.
\end{IEEEbiography}

\begin{IEEEbiography}[{\includegraphics[width=1in,height=1.25in,clip,keepaspectratio]{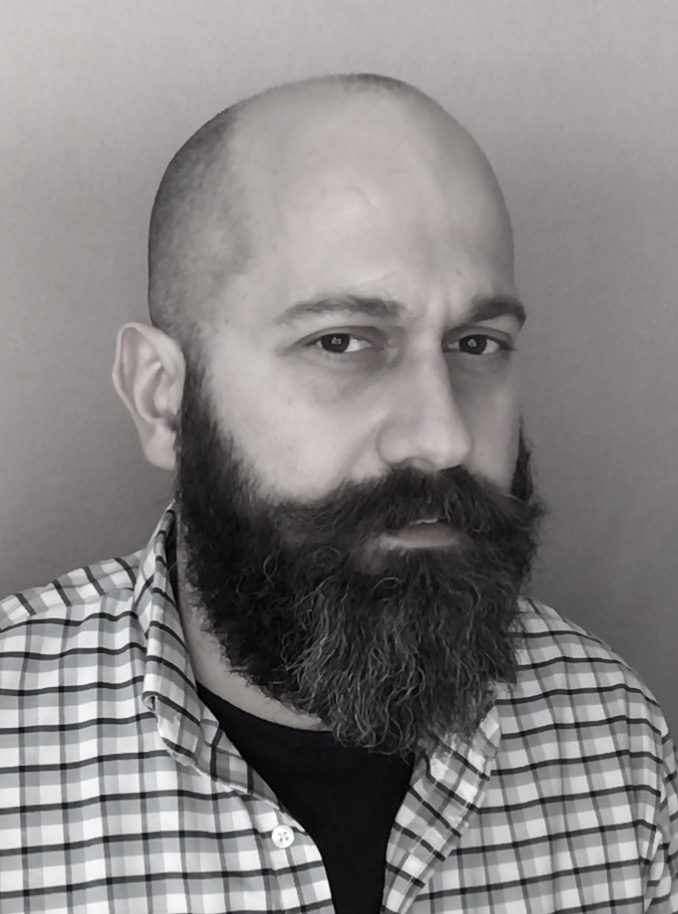}}]{George T Amariucai} was born and raised in Romania. He received his PhD in electrical and computer engineering from Louisiana State University (2009).

Amariucai is currently with the Department of Computer Science at Kansas State University. His research interests are focused on cyber security and its intersections with probability and information theory, applied and theoretical machine learning, wireless communication networks, cryptography, and social sciences. He is the director of the PITS Lab.
\end{IEEEbiography}

\EOD

\end{document}